\documentclass[superscriptaddress,twocolumn,aps,pra,showpacs,floatfix]{revtex4-1}
\usepackage{amsmath}
\usepackage{bm}
\usepackage{dcolumn}
\usepackage{graphicx}

\begin{document}

\title{Electron elastic scattering off a spin-polarized Cr atom}
\author{V. K. Dolmatov}
\affiliation{University of North Alabama,
Florence, Alabama 35632, USA}
\author{ M. Ya. Amusia}
\affiliation{Racah Institute of Physics, Hebrew University, 91904 Jerusalem, Israel}
\affiliation{A. F. Ioffe Physical-Technical Institute, 194021 St. Petersburg, Russia }
\author{L. V. Chernysheva}
\affiliation{A. F. Ioffe Physical-Technical Institute, 194021 St. Petersburg, Russia}
\date{\today}

\begin{abstract}
Electron elastic scattering off a spin-polarized Cr(...$3d^{5}4s^{1}$, $^{7}S$) atom is theoretically studied in the region of
electron energies up to $15$ eV using both  a one-electron ``spin-polarized''
Hartree-Fock and multielectron ``spin-polarized'' random phase approximation
with exchange. It is found that scattering phase shifts of oppositely
spin-polarized incoming electrons and corresponding cross sections of the
scattering reactions  significantly differ from each other, in general, even
without accounting for spin-orbit interaction. This is shown to be
associated with the presence of two semifilled $3d^{5}$ and $4s^{1}$ subshells in the Cr's configuration which induce
considerably different exchange in the interaction of oppositely
spin-polarized incoming electrons with the atom-target.
The importance of electron correlation in $e^{-}+\rm Cr$ elastic scattering
process is revealed. Moreover, correlation is shown to induce
strong differences between scattering of oppositely spin-polarized
electrons off Cr. A physically transparent interpretation for the latter is
provided.
\end{abstract}

\maketitle

\section{Introduction}

The Cr(...$3d^{5}4s^{1}$, $^{7}S$) atom is an attractive, even from
a purely theoretical point of view, object for studying of delicate exchange and
electron correlation interactions in electron-atom scattering. Indeed,
Cr is the first atom
in the periodic table with the maximum unbalanced number of spin-up
and spin-down electrons in its ground-state configuration, owing to its
two $3d^{5}$ and $4s^{1}$ semifilled subshells
with co-directed electron spin orientations. This
unbalance can lead to big differences in the exchange interaction between
incoming spin-up and spin-down electrons with the atom. It is of interest
and importance to study how this difference can affect electron scattering
phases and cross sections both from the viewpoint of pure static
exchange (e.g., at a Hartree-Fock approximation level) and electron
correlation. Whereas accounting for correlation in electron
scattering off atoms with arbitrary open shells encounters serious and not yet
entirely resolved difficulties, compared to scattering off closed shell
atoms, the situation is greatly simplified for the case of semifilled shell
atoms. This is because electrons with opposite spin projections can
be treated as two types of different particles, namely ``up'' (spin-up) and
``down'' (spin-down) electrons \cite{JETP83,Amusia}.
As a result, electron scattering off Cr inherits intrinsic properties of scattering off
both closed shell, on the one hand, and open shell, on the other hand, high
spin atoms. This bridges the way from relative ``simplicity'' in treating
closed shell atoms to utter ``complexity'' in doing so for the cohort of
open shell atoms. Furthermore, not the least important, owing to a not too
high nuclear charge of Cr, relativistic effects there can be discarded, to a
good approximation. This greatly simplifies the study of electron scattering
off Cr.

To the best of the authors' knowledge, there have been only two papers
dedicated to electron elastic scattering off Cr atom to date \cite{Hanne,Bartschat}.
Hanne et.\,al.\,\cite{Hanne} measured superelastic scattering of polarized electrons
from laser-excited polarized Cr$^{\ast }$ atoms at scattering angles ranging from $10^{\rm o}$ to $140^{\rm o}$
but only at two collision energies $\epsilon =6.8$ and $13.6$ eV.
The above experimental
study was followed by a theoretical work \cite{Bartschat}, where the
calculations were performed in the frame of a non-relativistic R-matrix
 approximation for the same differential scattering cross
section at the same two electron energies as in \cite{Hanne}. Thus, obviously,
the existing to date results on electron scattering off chromium provide only a limited
initial insight into the problem, because only a laser-excited Cr$^{\ast}$-target was
chosen and the scattering process itself was explored only at two collision energies.

We are not aware of any work on electron elastic scattering by the
ground-state Cr atom. Furthermore, clearly, studying the scattering process
through a continuum spectrum of electron energies is a way toward a deeper
understanding of electron-atom scattering, including the search for possible
shape and correlation resonances. Moreover, not the least important, the
study should highlight the role of unbalanced exchange in scattering of
spin-up and spin-down electrons off the Cr atom in a physically transparent
manner, to the benefit of a better understanding of the role of both
exchange and  correlation.

It is precisely the aim of this work to conduct the above outlined study in
order to get a deeper, clearer insight into electron elastic scattering off
Cr through a continuously changing range of electron energies. To meet the
goal, a ``spin-polarized'' Hartree-Fock approximation (SPHF) \cite{Slater,ATOM},
 a concept of the reducible self-energy part $\tilde{\Sigma}(\epsilon)$ of the
 Green function $G$ of an incoming electron, Feynman diagrammatic
technique as well as ``spin-polarized'' random phase approximation with
exchange (SPRPAE) \cite{ATOM} are combined together, as in \cite{DolAmCherPRA13},
in the performed study. The electron energy region of up
to $15$-$20$ eV, where most interesting effects occur, is considered.

Atomic units (a.u.)   are used throughout the paper
unless specified otherwise.

\section{Theory}

To learn about exchange effects in $e^{-}+\rm Cr$ elastic scattering initially
at the simplest static-exchange level, let us use the ideas of the SPHF
approximation. SPHF was originally suggested by Slater \cite{Slater} for
the calculation of the structure of semifilled shell atoms. Later,
it was extended to the calculation of photoionization of, as well as
electron scattering by, semifilled shell atoms, see
works \cite{JETP83,DolAmCherPRA13,JETP90,Remeta2010} and references therein. SPHF
accounts for the fact that spins of all electrons in semifilled subshells of
the ground-state atom - the $3d^{5}$ and $4s^{1}$
subshells in Cr - are co-directed, in accordance with Hund's rule. In the
present paper, for the sake of certainty, we consider spins of the the $3d$
and $4s$ electrons in the ground state of Cr being  directed upward
($\uparrow $). Then, each of the closed ${n\ell }^{2(2\ell +1)}$ subshells in
the atom splits into two semifilled subshells of opposite spin
orientations - ${n\ell }^{2\ell +1}$$\uparrow $ and ${n\ell }^{2\ell +1}$$%
\downarrow $ - whose electronic energies $\epsilon_{n\ell}$$\uparrow $
and $\epsilon_{n\ell }$$\downarrow $ as well as radial
wavefunctions $P_{\epsilon_{n\ell}}^{\uparrow }(r)$ and $P_{\epsilon_{n\ell}}^{\downarrow }(r)$
are different. The latter is because of the presence of
exchange interaction between corresponding ${n\ell}$$\uparrow$ spin-up electrons
with the $3d$$\uparrow $ and $4s$$\uparrow$
electrons from the half-filled subshells of the atom, but absence of such for ${n\ell}$$\downarrow$ electrons.
Correspondingly, in SPHF,
the Cr's ground-state configuration takes the following form:
Cr(...${3p}^{3}$$\uparrow $${3p}^{3}$$\downarrow $${3d}^{5}$$\uparrow $${4s}^{1}$$\uparrow $, $^{7}$S).
SPHF equations for the wavefunctions of the ground and excited states, as well as for the
wavefunctions $P_{\epsilon\ell}^{\uparrow}(r)$ and
$P_{\epsilon\ell}^{\downarrow}(r)$ of scattering states of Cr differ from ordinary HF equations for closed shell
atoms by accounting for exchange interaction only between electrons with the
same spin orientation ($\uparrow $, $\uparrow $ or $\downarrow $, $\downarrow $) \cite{ATOM}.
The $P_{\epsilon\ell}^{\uparrow}(r)$ and
$P_{\epsilon\ell}^{\downarrow}(r)$ functions have the standard central field
asymptotic behavior at large $r\gg 1$:

\begin{eqnarray}
P_{\epsilon \ell }^{\uparrow (\downarrow )}(r)\approx \frac{1}{\sqrt{\pi k}}%
\sin \left( kr-\frac{\pi \ell }{2}+\zeta _{\ell }^{\uparrow (\downarrow
)}(\epsilon )\right).
\end{eqnarray}
Here, $k$, $\ell $, and $\epsilon $ are the momentum, orbital momentum and
energy of a scattered electron, whereas $\zeta_{\ell}^{\uparrow(\downarrow)}(\epsilon)$
are elastic scattering phase shifts of the spin-up(down) electrons, respectively.
Corresponding total electron elastic
scattering cross sections $\sigma^{\uparrow}(\epsilon)$ and $\sigma^{\downarrow}(\epsilon)$
are then determined in the standard for a
spherical potential manner, as follows:

\begin{eqnarray}
\sigma ^{\uparrow(\downarrow)}(k)=\frac{4\pi}{k^{2}}\sum_{\ell
=0}^{\infty}(2\ell +1)\sin^{2}\zeta_{\ell}^{\uparrow (\downarrow)}(k).
\end{eqnarray}

In order to take into account the electron correlation in $e^{-}+\rm Cr$ elastic
scattering, let us exploit the concept of the so-called irreducible
self-energy part of the one-electron Green function $\Sigma(\epsilon)$ of a
scattering electron (see, e.\,g.\,, \cite{Amusia,ATOM,Abrikosov,Case_Stud} and references therein).
 In the present paper, the latter will be
accounted for in the framework of the random phase approximation with
exchange (RPAE) \cite{ATOM,DolAmCherPRA13}. Many-body RPAE uses the
Hartree-Fock atomic ground-state as the vacuum state. Alternatively, RPAE
theory, which chooses SPHF as the zeroth-order basis, is termed the
``spin-polarized'' RPAE (SPRPAE). SPRPAE is a straightforward modification of
$\Sigma(\epsilon )^{\rm RPAE}$ to the case of electron
scattering off semifilled shell atoms \cite{DolAmCherPRA13}.

\begin{figure}[tbp]
\includegraphics[width=\columnwidth]{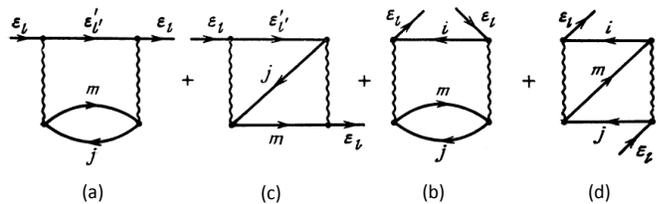}
\caption{The irreducible self-energy part $\tilde{\Sigma}^{\uparrow(\downarrow)}(\epsilon)$
of the Green function of a
spin-up(down) scattering electron in the SPRPAE$1$ approximation. Here,
a line with a right arrow denotes an electron, whether a scattered electron
(states $|\epsilon_{\ell}\rangle$ and $|\epsilon_{\ell^{\prime}}^{\prime}\rangle$) or an atomic excited electron (a state $|m \rangle$), a line with a
left arrow denotes a vacancy (hole) in the atom (states $\langle j|$ and $\langle i|$), a
wavy line denotes the Coulomb interelectron interaction $V$.}
\label{fig1}
\end{figure}

In the simplest second-order perturbation theory in the Coulomb
interelectron interaction $V$ between the incoming and atomic electrons, to
be labeled as SPRPAE$1$, the irreducible self-energy part of the
one-electron Green function of a spin-up [$\Sigma^{\uparrow}(\epsilon)$]
or spin-down [$\Sigma^{\downarrow}(\epsilon)$]
scattering electron is depicted with the help of Feynman diagrams in Fig.~\ref{fig1}.

The diagrams of Fig.~\ref{fig1} illustrate how a scattered electron ``$\epsilon _{\ell }$'' perturbs a $j$-subshell
of the atom by causing $j$ $\rightarrow $ $m$ excitations from the subshell and then gets coupled with
these excited states itself via the Coulomb interaction. Note that the
diagrams (c) and (d) are precisely due to exchange interaction between the
scattered  and atomic electrons. These diagrams are referred to as the
``exchange diagrams'' in the present paper. They vanish whenever spins of an
incoming electron and the perturbed atomic subshell have opposite directions, due to orthogonality of electron
spin-functions. Therefore, the number of diagrams that contribute to
scattering of spin-up electrons off Cr differs from the number of
corresponding diagrams for spin-down scattering electrons. This is because,
in the latter case, no exchange diagrams (c) and (d) exist when the perturbed
electrons are those from the semifilled $3d^{5}$$\uparrow$ or $4s^{1}$$\uparrow$
subshells. One thus immediately arrives at a physically transparent
conclusion that unbalanced correlation-exchange in $e^{-}+\rm Cr$ scattering
adds additional, compared to SPHF static-exchange, differences in scattering
of oppositely spin-polarized electrons off the Cr atom.

For a fuller account for electron correlation in $e^{-}+\rm Cr$ elastic scattering let us
introduce the reducible $\tilde{\Sigma}^{\uparrow(\downarrow)}(\epsilon)$ part of the self-energy part of
the one-electron Green function. The latter can be found as the solution of the following simplified  Dyson equation \cite{ATOM}, to a good approximation:
\begin{eqnarray}
\hat{\tilde{\Sigma}}^{\uparrow (\downarrow )}=\hat{\Sigma}^{(1)\uparrow
(\downarrow)}-\hat{\Sigma}^{(1)\uparrow (\downarrow)}\hat{G}^{(0)\uparrow
(\downarrow)}\hat{\tilde{\Sigma}}^{\uparrow(\downarrow)}
\label{Eq3}.
\end{eqnarray}
The Eq.~(\ref{Eq3}) is written in an operator form, where $\hat{\Sigma}^{(1)\uparrow (\downarrow)}$
is the operator of the  irreducible
self-energy part of the Green-function operator calculated in the
framework of SPRPAE$1$, $\hat{G}^{(0)\uparrow(\downarrow)}=(\hat{H}^{(0)\uparrow(\downarrow)}-\epsilon)^{-1}$ is the SPHF operator of the
electron's Green function, $\hat{H}^{(0)\uparrow(\downarrow)}$ is the SPHF
Hamiltonian operator of the electron-atom system. As in \cite{DolAmCherPRA13}, the approximation (\ref{Eq3}) will be referred to as SPRPAE$2$. Clearly,
it accounts for an infinite series of diagrams of Fig.~\ref{fig1} at
various combinations. Many blocks of these diagrams, namely, those which
contain exchange diagrams (c) and (d) of Fig.~\ref{fig1}, fall out from the
description of spin-down electron scattering. This is for the same reason
which was explained above in interpreting the significance of exchange
diagrams (c) and (d) of Fig.~\ref{fig1}. Thus, SPRPAE$2$ correlation will
additionally induce differences between scattering of spin-up and spin-down
electrons off Cr due to more complex spin-dependence of electron
correlation in the $e^{-}+\rm Cr$ system.

In the framework of SPRPAE$1$ or SPRPAE$2$, the elastic electron scattering
phase shifts $\zeta_{\ell}^{\uparrow(\downarrow)}$  are determined as \cite{ATOM}
\begin{eqnarray}
\zeta_{\ell }^{\uparrow(\downarrow)}=\delta_{\ell }^{\rm SPHF\uparrow(\downarrow)}+\Delta\delta_{\ell }^{\uparrow(\downarrow)}.
\end{eqnarray}
Here, $\Delta\delta_{\ell}^{\uparrow(\downarrow)}$ is the correlation
correction term to the SPHF calculated phase shift $\delta_{\ell}^{\rm SPHF\uparrow(\downarrow)}$:
\begin{eqnarray}
\Delta\delta_{\ell}^{\uparrow(\downarrow)}=\tan^{-1}\left(-\pi
\left\langle\epsilon\ell^{\uparrow(\downarrow)}|\tilde{\Sigma}^{\uparrow(\downarrow)}|\epsilon\ell^{\uparrow(\downarrow)}\right\rangle \right).
\end{eqnarray}
The mathematical expression for
$\left\langle\epsilon\ell^{\uparrow(\downarrow)}|\tilde{\Sigma}^{\uparrow(\downarrow)}|\epsilon\ell^{\uparrow(\downarrow)}\right\rangle$
is cumbersome. The
interested reader is referred to \cite{ATOM} for details. The matrix element
$\left\langle\epsilon\ell^{\uparrow(\downarrow)}|\tilde{\Sigma}^{\uparrow(\downarrow)}|\epsilon\ell^{\uparrow(\downarrow)}\right\rangle$ becomes
complex for electron energies exceeding the
ionization potential of the atom-target, and so does the correlation term
$\Delta\delta_{\ell}^{\uparrow(\downarrow)}$ and, thus, the phase shift
 $\zeta_{\ell}^{\uparrow(\downarrow)}$ as well. Correspondingly,
\begin{eqnarray}
\zeta_{\ell }^{\uparrow(\downarrow)}=\delta_{\ell}^{\uparrow(\downarrow)}+i\mu_{\ell}^{\uparrow(\downarrow)},
\end{eqnarray}
where
\begin{eqnarray}
\delta_{\ell}^{\uparrow(\downarrow)}=\delta_{\ell }^{\rm SPHF\uparrow(\downarrow)}+
\rm Re\Delta\delta_{\ell}^{\uparrow(\downarrow)},\quad \mu_{\ell}^{\uparrow(\downarrow)}=
Im\Delta\delta_{\ell }^{\uparrow(\downarrow)}.
\end{eqnarray}

The spin-up(down) total electron elastic scattering cross section $\sigma^{\uparrow(\downarrow)}$ is then given by the expression
\begin{eqnarray}
\sigma^{\uparrow (\downarrow)}=\sum_{\ell =0}^{\infty }\sigma_{\ell}^{\uparrow(\downarrow)},
\end{eqnarray}
where $\sigma_{\ell}^{\uparrow(\downarrow)}$ is the electron elastic
scattering partial cross section:
\begin{eqnarray}
\sigma_{\ell}^{\uparrow(\downarrow)}=\frac{2\pi }{k^2}(2\ell+1)\frac{\cosh{2\mu_{\ell}^{\uparrow(\downarrow)}}-
\cos{2\delta_{\ell}^{\uparrow(\downarrow)}}}{{\rm e}^{2\mu_{\ell}^{\uparrow(\downarrow)}}}.
\label{Eq9}
\end{eqnarray}

\section{Results and discussion}

\subsection{$e^{-}+\rm Cr$ elastic scattering phase shifts $\zeta_{\ell}(\epsilon)$}

A calculation shows that, at electron energies of up to $20$ eV, it is
sufficient to account for the contributions of only $s$, $p$, $d$ and $f$ partial waves to the total elastic
scattering cross section. Likewise, accounting for only monopole, dipole,
quadrupole and octupole excitations from the $4s^{1}$$\uparrow $
and $3d^{5}$$\uparrow $ subshells of Cr in calculations of
scattering phase shifts is a good approximation as well.

\begin{figure}[h]
\centering \includegraphics[width=\columnwidth]{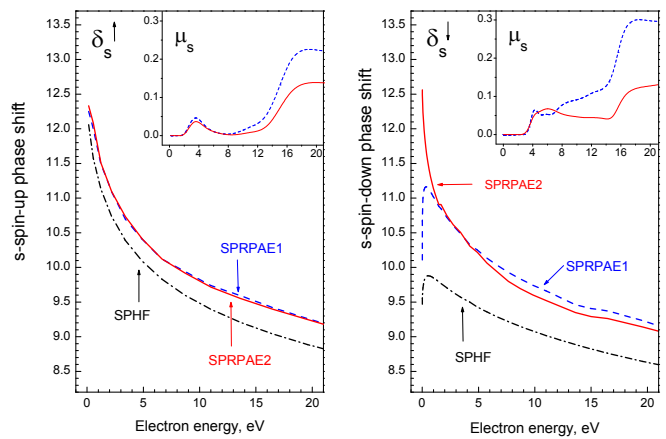}
\caption{(Color online) SPHF, SPRPAE$1$ and SPRPAE$2$ calculated (in
units of radians) real $\delta_{s}^{\uparrow}$ and
imaginary $\mu_{s}^{\uparrow}$ parts of the $s$$\uparrow$-phase shift (left panel)
as well as those $\delta_{s}^{\downarrow}$ and $\mu_{s}^{\downarrow}$
of the $s$$\downarrow$-phase shift (right panel) of $e^{-}+\rm Cr$ elastic
scattering, as marked.}
\label{fig2}
\end{figure}

Corresponding SPHF,
SPRPAE$1$ and SPRPAE$2$ calculated real $\delta_{s}^{\downarrow}(\epsilon)$
and imaginary $\mu_{s}^{\downarrow}(\epsilon)$ parts of
elastic scattering phase shifts $\zeta_{s}^{\downarrow}(\epsilon)$ and $\zeta_{s}^{\uparrow}(\epsilon)$ of
$s$-spin-down and
$s$-spin-up electronic waves are depicted in Fig.~\ref{fig2}.

First, note how different are the SPHF calculated $\delta_{s}^{\downarrow}(\epsilon)$ and
$\delta_{s}^{\uparrow}(\epsilon)$ at $\epsilon $ $\rightarrow$ $0$.
This finds a ready explanation on the basis of Levinson's
theorem \cite{Landau} which we write as
\begin{eqnarray}
\left. \delta_{\ell}(\epsilon)\right\vert_{\epsilon \rightarrow 0}\rightarrow (n_{\ell }+q_{\ell })\pi.
\end{eqnarray}
Here $q_{\ell}$ is the number of electronic subshells with given $\ell$ in
the ground-state configuration of the atom, whereas $n_{\ell }$ is the
number of additional bound states with the same $\ell$ in the very field of
the neutral atom. As known, the HF field of a neutral atom cannot bind
an additional electron. Correspondingly, as $\epsilon \rightarrow 0$,
the phase shift $\delta_{s}^{\uparrow}(\epsilon)$ $\rightarrow $ $4\pi$, since
there are four $ns$$\uparrow$ subshells in the ground-state
of Cr, whereas $\delta_{s}^{\downarrow}(\epsilon)$ $\rightarrow $ $3\pi$,
in view of the absence of $ns$$\downarrow$ bound states with $n\geq 4$ in the atom.
Thus, the calculation reveals that the inherent to Cr
different exchange between spin-up and spin-down incoming electrons
with the atomic electrons is significant already in the one-electron SPHF approximation.

Next, note how generally stronger electron correlation affects the s-phase
shift of spin-down electrons than that of spin-up electrons over the whole
range of considered energies. This provides a numerical support to the
suggested in the previous section possibility for a noticeable spin-dependence of
correlation in the $e^{-}+\rm Cr$ system.

Furthermore, note that SPRPAE$2$ correlation affects scattering of spin-down electrons
stronger than does lower-order SPRPAE$1$ correlation. This is in contrast to scattering of spin-up
electrons, where SPRPAE$1$ and SPRPAE$2$ calculated corrections to $\delta_{s}^{\uparrow}(\epsilon)$
 are practically identical.
In particular, SPRPAE$2$ correlation radically, both quantitatively and
qualitatively, changes the behavior of $\delta_{s}^{\downarrow}(\epsilon)$
at $\epsilon$ $\rightarrow$ $0$ compared to SPHF or SPRPAE$1$ calculated
data. As a result, SPRPAE$2$ calculated $\delta_{s}^{\downarrow}(\epsilon)$ approaches $4\pi$
rather than $3\pi$ (as in SPHF and SPRPAE$1$ calculations) at $\epsilon$ $\rightarrow$ $0$.
With Levinson's theorem in mind, this is indicative of the emergence of a
correlation induced additional $ns$$\downarrow$ bound state with $n\geq 4$ in the field of Cr,
i.e., of the formation of a negative chromium ion, possibly of
Cr$^{-}$(...$3p^{3}$$\uparrow$$3p^{3}$$\downarrow$$3d^{5}$$\uparrow$$4s$$\uparrow$$4s$$\downarrow$, $^{6}$S).
As known, see, e.g., \cite{Cr-}, the Cr$^{-}$(...$3d^{5}4s^{2}$, $^{7}S)$ anion does indeed exist in nature.
Clearly, the ability of SPRPAE$2$ to re-discover the existence of Cr$^{-}$ speaks in favor of this approximation.
In addition, it underpins, once again, the importance, as well as spin-dependent specificity,
of electron correlation in the $e^{-}+\rm Cr$ system.

\begin{figure}[h]
\centering
\includegraphics[width=\columnwidth]{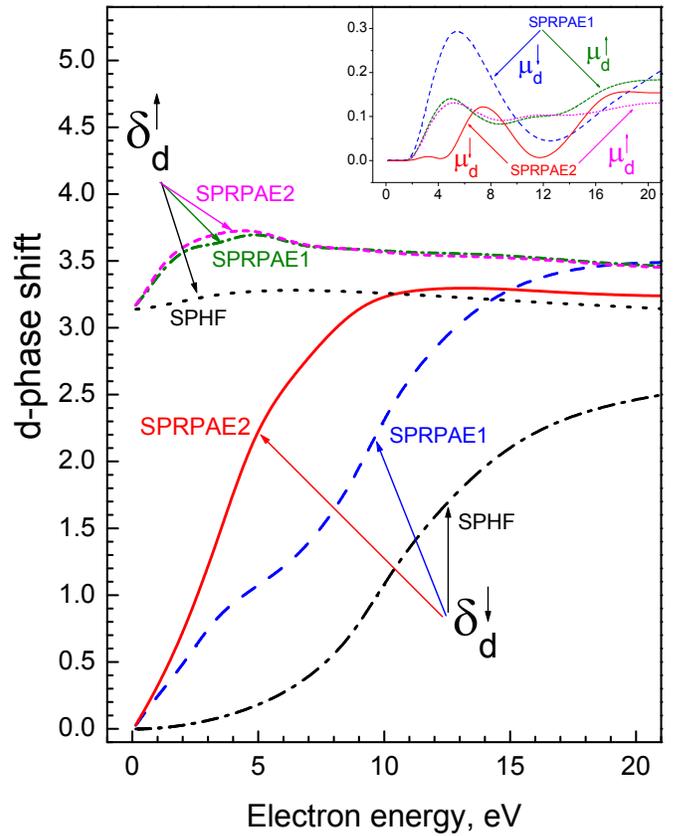}
\caption{(Color online) SPHF, SPRPAE$1$ and SPRPAE$2$ calculated data (in
units of radians) for $e^{-}+\rm Cr$ real $\delta_{d}^{\uparrow(\downarrow)}$ (main panel) and
 imaginary $\mu_{d}^{\uparrow(\downarrow)}$ (inset) parts of the scattering phase shifts
 $\zeta_{d}^{\uparrow(\downarrow)}$ of $\epsilon d$$\uparrow$ and $\epsilon d$$\downarrow$ partial electronic waves, as marked.}
\label{fig3}
\end{figure}

Impressive differences also emerge between calculated electron elastic
scattering phase shifts of spin-up and spin-down $\epsilon d$-electron
waves. Corresponding $\zeta_{d}^{\uparrow}(\epsilon)$ and $\zeta_{d}^{\downarrow}(\epsilon)$ are depicted in Fig.~\ref{fig3}.

First, note that $\delta_{d}^{\uparrow}(\epsilon)$ $\rightarrow$ $\pi$ whereas $\delta_{d}^{\downarrow}(\epsilon)$ $\rightarrow$ $0$
at $\epsilon $ $\rightarrow $ $0$ in each of the three approximations.
One concludes that, contrary to the $s$-wave,
calculated results demonstrate the incapability of neutral Cr to capture and bind an additional $nd$$\uparrow$ or $nd$$\downarrow$ electron.
Indeed, in this case, the parameters of Levinson's theorem take the following values: $n_{d\uparrow}=n_{d\downarrow}=0$,
but $q_{d\uparrow}=1$,  because of the presence of $3d^{5}$$\uparrow$ subshell in
the Cr structure,  whereas $q_{d\downarrow}=0$,  since there is no $nd$$\downarrow$ bound states in the ground-state of the atom.
Consequently, $\delta_{d}^{\uparrow}(\epsilon)$ $\rightarrow$ $\pi$,
whereas $\delta_{d}^{\downarrow}(\epsilon)$ $\rightarrow$ $0$ at $\epsilon$~$\rightarrow$~$0$,
exactly as was obtained in the above commented calculation. The fact that calculated
$\delta_{d}^{\uparrow}(\epsilon)$ and $\delta_{d}^{\downarrow}(\epsilon)$ are in agreement with Levinson's theorem
speaks, once again, to the capability of the used theory to reveal differences between
scattering of oppositely spin-polarized electrons off Cr atom.

Second, as in the case of the $s$-scattering, note the
significance of electron correlation in determining the $\delta_{d}^{\uparrow\downarrow}$ phase shifts.
Most importantly, however, is that
correlation effects are considerably stronger in $\delta_{d}^{\downarrow}$ than in $\delta_{d}^{\uparrow}$.
This, once again, underlines a great role
played by unbalanced correlation-exchange in scattering of oppositely
spin-polarized electrons off Cr.

\begin{figure}[h]
\centering \includegraphics[width=\columnwidth]{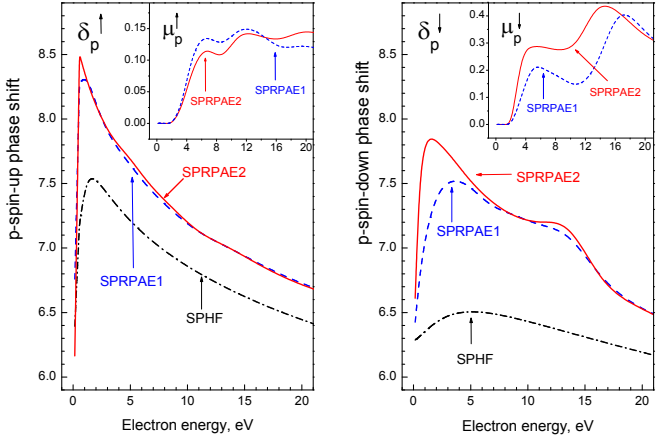}
\caption{(Color online) SPHF, SPRPAE$1$ and SPRPAE$2$ calculated data (in
units of radians) for real $\delta_{p}^{\uparrow(\downarrow)}$ and imaginary
$\mu_{p}^{\uparrow(\downarrow)}$
parts of the $e^{-}+\rm Cr$ elastic scattering phase shifts $\zeta_{p}^{\uparrow(\downarrow)}$, as marked.}
\label{fig4}
\end{figure}

\begin{figure}[h]
\centering \includegraphics[width=\columnwidth]{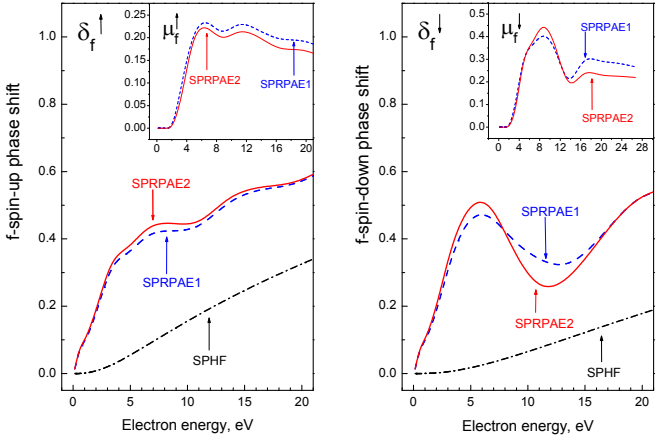}
\caption{(Color online) SPHF, SPRPAE$1$ and SPRPAE$2$ calculated data (in
units of radians) for real $\delta_{f}^{\uparrow\downarrow}$
and imaginary $\mu_{f}^{\uparrow\downarrow}$ parts of the $e^{-}+\rm Cr$ elastic scattering phase shifts $\zeta_{f}^{\uparrow(\downarrow)}$, as marked.}
\label{fig5}
\end{figure}

Calculated data for $e^{-}+\rm Cr$ scattering phase shifts $\zeta_{p}^{\uparrow(\downarrow)}(\epsilon)$
and $\zeta_{f}^{\uparrow(\downarrow)(\epsilon)}$
are depicted in Figs.~\ref{fig4} and \ref{fig5}, respectively.

Similarly to the previously discussed cases, results of the calculation reveal noticeable differences between $\zeta_{p}^{\uparrow}$ and $\zeta_{p}^{\downarrow}$,
as well as between $\zeta_{f}^{\uparrow}$ and $\zeta_{f}^{\downarrow}$ phase shifts, especially the differences brought about by
correlation. Once again, a much stronger correlation impact on scattering of
spin-down electrons than on scattering of spin-up electrons off Cr is uncovered.

\subsection{$e^{-}+\rm Cr$ elastic scattering cross sections}

\begin{figure}[h]
\centering \includegraphics[width=\columnwidth]{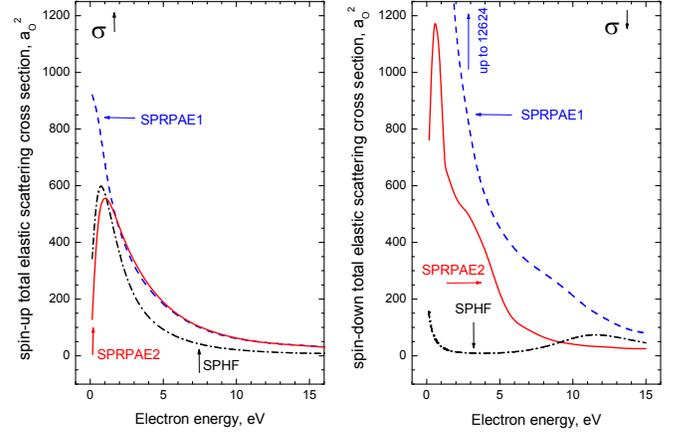}
\caption{(Color online) SPHF, SPRPAE$1$ and SPRPAE$2$ calculated $e^{-}+\rm Cr$
total elastic scattering cross sections $\sigma^{\uparrow}(\epsilon)$ and $\sigma^{\downarrow}(\epsilon)$ for spin-up
and spin-down scattering electrons, as marked.}
\label{fig6}
\end{figure}

Calculated data for the $e^{-}+\rm Cr$ total spin-up $\sigma^{\uparrow}(\epsilon)$ and
spin-down $\sigma^{\downarrow}(\epsilon)$ elastic scattering cross sections are
depicted in Fig.~\ref{fig6}.

Note large differences, both quantitative and qualitative, between $\sigma^{\uparrow }(\epsilon)$ and $\sigma^{\downarrow }(\epsilon)$ calculated
in each of the three approximations. Furthermore, note how stronger
correlation effects are in $\sigma^{\downarrow}(\epsilon)$ than in $\sigma^{\uparrow}(\epsilon)$. Moreover, note that, once again,
a fuller account of correlation in the framework of SPRPAE$2$ is much more important in the calculation of scattering of spin-down than
spin-up electrons off Cr.

\begin{figure}[h]
\centering \includegraphics[width=\columnwidth]{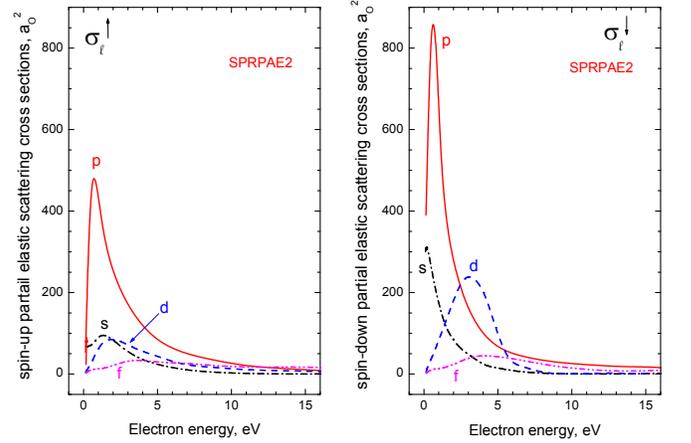}
\caption{(Color online) SPRPAE$2$ calculated $e^{-}+\rm Cr$ partial elastic
scattering cross sections $\sigma_{\ell}^{\uparrow(\downarrow)}$, as
marked.}
\label{fig7}
\end{figure}

In order to better understand details of the SPRPAE$2$ calculated total electron scattering cross section $\sigma^{\downarrow}(\epsilon)$ and $\sigma^{\uparrow}(\epsilon)$,
let us explore
corresponding SPRPAE$2$ calculated partial cross sections $\sigma_{\ell}^{\uparrow}(\epsilon)$ and $\sigma_{\ell}^{\downarrow}(\epsilon)$. The latter two are plotted in Fig.~\ref{fig7}.

By analyzing Figs.~\ref{fig6} and \ref{fig7}, one can see that a sharp
narrow maximum in $\sigma^{\rm sprpae2\downarrow}(\epsilon)$ at about $0.5$ eV
is due primarily to scattering of a $\epsilon p$$\downarrow$-electronic wave, whereas a shallow
maximum between approximately $2$ and $5$ eV is
brought about by  a $\epsilon d$$\downarrow$ scattering
wave. Similarly, the maximum in $\sigma^{\rm sprpae2\uparrow}$ at about $1.2$ eV is
 primarily due to scattering of a $\epsilon p$$\uparrow$-electronic wave. Note the absence of
a second shallow maximum in  $\sigma^{\rm sprpae2\uparrow}$, which would be similar to that in $\sigma^{\rm sprpae2\downarrow}(\epsilon)$ between $2$ and $5$ eV.
This is because the phase shift $\delta_{d}^{\downarrow}(\epsilon)$ passes
 through $\delta_{d}^{\downarrow}(\epsilon)=\pi/2$
with decreasing energy $\epsilon$, thereby causing the maximum in partial $\sigma_{d}^{\downarrow}(\epsilon)$, but $\delta_{d}^{\uparrow}(\epsilon)$
does not. Also note that plotted
in Fig.~\ref{fig7} data provide the evidence for sufficiency of accounting
for only $s$, $p$, $d$, and $f$ partial
waves in the calculation of $e^{-}+\rm Cr$ elastic scattering, in the considered
region of electron energies.

\section{Conclusion}

In conclusion, the performed study results in a deeper understanding of
electron elastic scattering off Cr, reveals novel interesting features
emerging in the scattering process due to unbalanced exchange in the atomic
system, as well as provides a physically transparent interpretation for the unraveled
effects at a one-electron and multielectron levels of sophistication.
Furthermore, this paper adds novel concrete data on $e+\rm Cr$ elastic scattering
to the existing database of physics quantities.

\section{Acknowledgements}

V.K.D. acknowledges the support of NSF Grant No.\ PHY-$1305085$.

\end{document}